\documentclass{llncs}
\usepackage{graphicx}
\usepackage{amsmath}
\usepackage{amssymb}
\usepackage{enumerate}
\usepackage{varioref}

\def\boxit#1{\vbox{\hrule\hbox{\vrule\kern4pt
  \vbox{\kern1pt#1\kern1pt}
\kern2pt\vrule}\hrule}}

\usepackage{url}
\newcommand\nc{\newcommand}

\newtheorem{rules}{Rule}

\nc{\crl}[2]{\begin{corollary}\label{crl:#1} #2 \end{corollary}}
\nc{\dfn}[2]{\begin{definition}\label{def:#1} #2 \end{definition}}
\nc{\lem}[2]{\begin{lemma}\label{lem:#1} #2 \end{lemma}}
\nc{\prp}[2]{\begin{proposition}\label{prp:#1} #2
\end{proposition}}
\nc{\thm}[2]{\begin{theorem}\label{thm:#1} #2\end{theorem}}
\nc{\fac}[2]{\begin{lemma}\label{fact:#1} #2 \end{lemma}}
\nc{\rul}[2]{\begin{rules}\label{rul:#1} #2 \end{rules}}

\nc{\eqn}[2]{\begin{eqnarray}\label{eqn:#1} #2 \end{eqnarray}}
\nc{\fig}[4]{\begin{figure}[h]
\begin{center}
\includegraphics[width=#2\textwidth]{#4}
\end{center}
\caption{#3}\label{fig:#1}
\end{figure}}

\nc{\tbl}[3]{\begin{table}[hbt] #3 \caption{#2} \label{tab:#1}
\end{table}}

\nc{\refc}[1]{Corollary~\ref{crl:#1}}
\nc{\refd}[1]{Definition~\ref{def:#1}}
\nc{\reff}[1]{Figure~\ref{fig:#1}}
\nc{\refl}[1]{Lemma~\ref{lem:#1}}
\nc{\refp}[1]{Proposition~\ref{prp:#1}}
\nc{\reft}[1]{Theorem~\ref{thm:#1}} \nc{\refe}[1]{(\ref{eqn:#1})}
\nc{\reftb}[1]{Table~\ref{tab:#1}}

\nc{\reffc}[1]{Fact~\ref{fact:#1}}
\nc{\refr}[1]{Rule~\ref{rul:#1}}

\nc{\pf}[1]{ \noindent
{\bf Proof.}  #1 \hfill \qed\par}

\long\def\invis#1{}

\title{A Parameterized Algorithm for Bounded-Degree Vertex Deletion}
\author{Mingyu Xiao\thanks{
Supported by NFSC of China under the Grant
61370071 and Fundamental Research Funds for the Central Universities under the Grant ZYGX2015J057.
}
}

 \institute{
 School of Computer Science and Engineering,\\
University of Electronic Science and Technology of China, China\\
 \email{myxiao@gmail.com}
}

\begin{document}

\maketitle

\begin{abstract}
The $d$-bounded-degree vertex deletion problem, to delete at most $k$ vertices in a given graph to make
the maximum degree of the remaining graph at most $d$, finds applications in computational biology, social network analysis and some others. It can be regarded as a special case of the $(d+2)$-hitting set problem and generates the famous vertex cover problem. The $d$-bounded-degree vertex deletion problem is NP-hard for each fixed $d\geq 0$.
In terms of parameterized complexity, the problem parameterized by $k$ is W[2]-hard for unbounded $d$ and fixed-parameter tractable for each fixed $d\geq 0$.
Previously, (randomized) parameterized algorithms for this problem with running time bound $O^*((d+1)^k)$ are only known for $d\leq2$.
In this paper, we give a uniform parameterized algorithm deterministically solving this problem in $O^*((d+1)^k)$ time for each $d\geq 3$.
Note that it is an open problem whether the $d'$-hitting set problem can be solved in $O^*((d'-1)^k)$ time for $d'\geq 3$.
Our result answers this challenging open problem affirmatively for a special case.
Furthermore, our algorithm also gets a running time bound of $O^*(3.0645^k)$ for the case that $d=2$, improving the previous deterministic bound of $O^*(3.24^k)$.

 \vspace*{5mm} \noindent {\bf Key words.} \ \
Parameterized algorithms, Graph algorithms, Bounded-degree vertex deletion, Hitting set
\end{abstract}

\section{Introduction}
%%%%%%%%%%%%%%%%%%%%%%
The $d$-bounded-degree vertex deletion problem is a natural generation of the famous vertex cover problem, which is one of the best studied problems in combinatorial optimization.
An application of the $d$-bounded-degree vertex deletion problem  in computational biology is addressed by Fellows et. al.~\cite{FG:gNT}: A clique-centric approach in the analysis of genetic networks based on micro-array data
can be modeled as the $d$-bounded-degree vertex deletion problem.
The problem also plays an important role in the area of property testing~\cite{NS:test}.
Its ``dual problem''-- the $s$-plex problem was introduced
in 1978 by Seidman and Foster~\cite{SF:plex} and it becomes an important problem in social network analysis now~\cite{BBH:plex}.

The $d$-bounded-degree vertex deletion problem is also extensively studied in theory, especially in parameterized complexity.
It has been shown that the problem parameterized by the size $k$ of the deletion set is W[2]-hard for unbounded $d$ and fixed-parameter tractable for each fixed $d\geq 0$~\cite{FG:gNT}.
Betzler et. al.~\cite{BBNU:treewidth} also studied the parameterized complexity of the problem with respect to  the treewidth $tw$ of the graph. The problem is FPT with parameters $k$
and $tw$ and W[2]-hard with only parameter $tw$.
Fellows et. al.~\cite{FG:gNT} generated the NT-theorem for the vertex cover problem to the $d$-bounded-degree vertex deletion problem, which can imply a linear vertex kernel for
the problem with $d=0, 1$ and a polynomial vertex kernel for each fixed $d\geq 2$. A linear vertex kernel for the case that $d=2$ was developed in~\cite{CF:copath}.
Recently,  a refined generation of the NT-theorem was proved~\cite{X:gNT}, which can get a linear vertex kernel for each fixed $d\geq 0$.

In terms of parameterized algorithms, the case that $d=0$, i.e., the vertex cover problem,  can be solved in $O^*(1.2738^k)$ time now~\cite{CKX:VC}.
When $d=1$, the problem is  known as the $P_3$ vertex cover problem. Tu~\cite{tu:p3vc} gave an $O^*(2^k)$-time algorithm and the running time bound was improved to $O^*(1.882^k)$ by Wu~\cite{wu:p3} and to $O^*(1.8172^k)$ by Katreni\v{c}~\cite{Katrenic:3PVC}.
When $d=2$, the problem is  known as the co-path/cycle problem. For this problem, there is an $O^*(3.24^k)$-time deterministic algorithm~\cite{CF:copath} and an $O^*(3^k)$-time randomized algorithm~\cite{feng:co-path}.
 For $d\geq 3$, a simple branch-and-reduce algorithm that tries all $d+2$ possibilities for a $(d+1)$-star in the graph gets the running time bound of $O^*((d+2)^k)$.
%We do not find any other special parameterized algorithms for this problem with $d\geq 3$.
In fact, the $d$-bounded-degree vertex deletion problem can be regarded as a special case of
the $(d+2)$-hitting set problem and the latter problem has been extensively studied in parameterized algorithms~\cite{NR:3HS,fernau:3HS,fernau:WHS,fernau:HS}.
For a graph $G$, we regard each vertex in the graph as an element and each $(d+1)$-star as a set of size $d+2$
(a vertex of degree $d_0>d$ will form $d_0 \choose d+1$ sets). Then the $d$-bounded-degree vertex deletion problem in $G$ becomes an instance of the $(d+2)$-hitting set problem.
There are several parameterized algorithms for the $d'$-hitting set problem running in $O^*((d'-1+c)^k)$ time~\cite{NR:3HS,fernau:HS}, where $0<c<1$ is a function of $d'^{-1}$.
It leaves as an interesting open problem whether the $d'$-hitting set problem can be solved in $O^*((d'-1)^k)$ time. Note that it is marked in \cite{fernau:HS} that ``$(d'-1)^k$ seems an unsurpassable lower bound''. By using fastest algorithms for the $(d+2)$-hitting set problem, we can
get an algorithm with running time bound of $O^*((d+1+c_0)^k)$ with $0<c_0<1$ for each fixed $d$.

In this paper, we design a uniform algorithm for the $d$-bounded-degree vertex deletion problem, which achieves the running time bound of $O^*((d+1)^k)$ for each $d\geq 3$.
Although our problem is a special case of the $(d+2)$-hitting set problem, the above bound is not easy to reach. We need a very careful analysis and some good graph structural properties.
It is also worthy to mention that our algorithm also works on the case that $d=2$ and runs in $O^*(3.0645^k)$ time, improving the previous deterministic bound of  $O^*(3.24^k)$~\cite{CF:copath} and comparable with the previous
randomized bound of $O^*(3^k)$~\cite{feng:co-path}.

\section{Preliminaries}
%%%%%%%%%%%%%%%%%%%%%%%%%%%%%%%%%%%%%%
\label{sec_Pre}

Let $G=(V,E)$ be a simple undirected graph, and $X\subseteq V$ be a subset of vertices.
The subgraph induced by $X$ is denoted by $G[X]$, and $G[V\setminus X]$ is written as $G\setminus X$.
We may simply use $v$ to denote the set $\{v\}$ of a single vertex $v$.
Let
$N(X)$ denote the set of {\em neighbors} of $X$, i.e.,
the vertices in $V\setminus X$ adjacent to a vertex $x\in X$,
and denote $N(X)\cup X$ by $N[X]$.
The \emph{degree} $d(v)$ of a vertex $v$ is defined to be $|N(v)|$.
A graph of maximum degree $p$ is also called a \emph{degree-$p$ graph}.
For  an integer $q\geq 1$, a star with $q+1$ vertices is called a  {\em $q$-star}.
A set $S$ of vertices is called a \emph{$d$-deletion set} of a graph $G$, if $G\setminus S$ has maximum degree at most $d$.
In our problem, we want to find a $d$-deletion set of size at most $k$  in a graph. Formally, our problem is defined as following.

\noindent\rule{\linewidth}{0.2mm}
\textsc{$d$-Bounded-Degree Vertex Deletion}\\
\textbf{Instance:} A graph $G=(V,E)$ and two nonnegative integers $d$ and $k$. \\
%\textbf{Parameter:} $k$.\\
\textbf{Question:} To decide whether there is a subset $S\subseteq V$ of vertices such that $|S| \leq k$
and the induced graph $G[V\setminus S]$ has maximum degree at most $d$.\\
\rule{\linewidth}{0.2mm}
%\medskip
In the above definition, $S$ is also called a \emph{solution set}.

\subsection{Some basic properties}

The following lemmas are basic structural properties used to design branching rules in our algorithms.

\lem{base1}{Let $v$ be a vertex of degree $\geq d+1$ in a graph $G$. Any $d$-deletion set contains either $v$ or $d(v)-d$ neighbors of $v$.
}

A vertex $v$ \emph{dominates} a vertex $u$ if all vertices of degree $\geq d+1$ in $N[u]$ are also in $N[v]$. Note that in this definition, we do not require $N[u]\subseteq N[v]$.

\lem{domination}{If a vertex $v$ of degree $d+1$ dominates a neighbor $u$ of it, then there is a minimum $d$-deletion set containing at least one vertex in $N[v]\setminus \{u\}$.
}
\pf{Since $v$ is of degree $d+1$, any $d$-deletion set $S$ contains at least one vertex in $N[v]$.
Assume that $S$ contains only $u$ in $N[v]$. We can see that $S'=S\cup\{v\}\setminus \{u\}$ is still a  $d$-deletion set and $|S'|\leq |S|$.
Thus, the lemma holds.
}

\lem{domination2}{If a vertex $u$ dominates a vertex $v$ of degree $d+1$, then there is a minimum $d$-deletion set containing at least one neighbor of $v$.
}
\pf{ Since $u$ dominates $v$ and $v$ is of degree $d+1$, we know that $u$ is a neighbor of $v$.
Any $d$-deletion set $S$ contains at least one vertex in $N[v]$ since it is of degree $d+1$.
Assume that $S\cap N[v]=\{v\}$. We can see that $S'=S\cup\{u\}\setminus \{v\}$ is a  $d$-deletion set
containing a neighbor of $v$ and $|S'|\leq |S|$.
Thus, the lemma holds.
}

%Note that the above two lemmas imply that in some cases there exists a minimum $d$-deletion set not containing a dominated vertex $u$.

If there is a vertex of degree $\geq d+1$ dominating a neighbor of it or being dominated by another vertex, we say that the graph has a \emph{proper domination}.
Note that if a vertex $u$ of degree $\geq d+1$ has at most one neighbor $v$ of degree $\geq d+1$, then $u$ is dominated by $v$ and then there is a proper domination. In fact, we have:
\lem{domi3}{
If a graph has no proper domination, then each vertex of degree $\geq d+1$ in it has at least two nonadjacent neighbors of
degree $\geq d+1$.}

\subsection{Branch-and-search algorithms}
Our algorithm is a typical branch-and-search algorithm.
In our algorithm, we search a solution for an instance by recursively
 branching on the current
instance into several smaller instances until the instances become trivial instances.
Each simple branching operation creates a recurrence relation.
Assume that the branching operation branches on an instance with parameter $k$  into $l$ branches such that
in the $i$-th branch the parameter decreases by
at least $a_i$. Let $C(k)$ denote the worst size of the search tree to search a solution to any instance with parameter $k$. We get a recurrence relation \footnote{In fact, we may simply write a recurrence relation as $C(k)\leq  C(k-a_1)+C(k-a_2)+\cdots +C(k-a_l)$. This difference will only affect a constant behind $O$ in the finial running time.}
$$C(k)\leq  C(k-a_1)+C(k-a_2)+\cdots +C(k-a_l)+1.$$
The largest
root of the function $f(x)=1-\sum_{i=1}^l x^{-a_i}$ is called the \emph{branching factor} of the recurrence relation.
Let $\alpha$ be the maximum branching factor among all branching factors in the algorithm.
The size of the search tree that represents the branching process of the algorithm applied to
an instance with parameter $k$ is  given by $O(\alpha^k)$.
More details about the analysis and how to solve recurrences
can be found in the monograph~\cite{Fomin:book}.
%Note that for two recurrences $C(n)\leq  \sum_{i=1}^lC(n-a_i)$ and $C(n)\leq \sum_{i=1}^lC(n-b_i)$
%with $a_i\geq b_i$ $(i=1,2,\dots,l)$,
%the branch factor of the first recurrence is not greater than this of the second one.

\section{The idea and organization of the algorithm}
Our purpose is to design a branch-and-search algorithm for the $d$-bounded-degree vertex deletion problem such
that the branching factor of each recurrence relation with respective to the parameter $k$ is at most $d+1$.
\refl{base1} provides a simple branching rule: for a vertex $v$ of degree $\geq d+1$, branching
by either  including $v$ or each set of $d(v)-d$ neighbors of $v$ to the solution set.
We will show that when $d(v)\geq d+2$, this simple branching operation is good enough to get a branching factor $\leq d+1$ for each $d\geq 2$ (See Step 1 in Section~\ref{sec_alg}). Thus, we can use this operation to deal with vertices of degree $\geq d+2$.
\refl{base1} for a degree-($d+1$) vertex $v$ can be interpreted as: at least one vertex in $N[v]$ is in a $d$-deletion set. This branching operation will only get a branching factor of $d+2$ for this case.
But when there is a proper domination in a degree-($d+1$) graph, we still can branch with branching factor $d+1$, since
we can ignore one branch by \refl{domination} and \refl{domination2}. The detailed analysis is given in Step 2 in Section~\ref{sec_alg}.
When the graph is of maximum degree $d+1$ and has no proper domination, we need to use more structural properties.

%Compared the $d$-bounded-degree vertex deletion problem in graphs of maximum degree $d+1$ and the $(d+2)$-hitting set problem
%Note that for two adjacent degree-$(d+1)$ vertices $v_1$ and $v_2$, there are at least two vertices in the intersection
%If two degree-$(d+1)$ vertices $v_1$ and $v_2$ have some common neighbors, then the intersection of $N(v_1)$
% and $N(v_2)$ is not an empty set.

To find a $d$-deletion set in a degree-$(d+1)$ graph is equivalent to find a vertex subset intersecting $N[v]$
for each degree-($d+1$) vertex $v$.
If there are some vertices in $N[v_1]\cap N[v_2]$ for two degree-$(d+1)$ vertices $v_1$ and $v_2$, some information
may be useful for us to design a good branching rule.
Note that for two adjacent degree-$(d+1)$ vertices $v_1$ and $v_2$, there are at least two vertices in the intersection
of $N[v_1]$ and $N[v_2]$. \refl{domi3} guarantees that each degree-$(d+1)$ vertex has at least two nonadjacent degree-$(d+1)$ neighbors if a degree-$(d+1)$ graph has no proper domination.
So we will focus on adjacent degree-$(d+1)$ vertices.

We define three relations between two degree-($d+1$) vertices.  A pair of adjacent degree-($d+1$) vertices is a \emph{good pair} if they have at least one
and at most $d-2$ common neighbors. A pair of adjacent degree-($d+1$) vertices is a \emph{close pair} if they have exactly $d-1$ common neighbors.
A pair of nonadjacent degree-($d+1$) vertices is a \emph{similar pair} if they have the same neighbor set.
We have a good branching rule to deal with good pairs. See Step 3 in Section~\ref{sec_alg}.
After dealing with all good pairs, for any pair of adjacent degree-($d+1$) vertices, either it is a close pair or the two vertices have no common neighbor.
We do not have a simple branching rule with branching factor $d+1$ for these two cases.
Then we change to consider three adjacent  degree-($d+1$) vertices.

Let $v_1, v_2$ and $v_3$ be three degree-($d+1$) vertices such that $v_2$ is adjacent to $v_1$ and $v_3$. We find that the hardest case is that exact one pair of vertices in $\{v_1,v_2,v_3\}$
is a close or similar pair, for which we still can not get a branching factor $\leq d+1$. We call this case a \emph{bad case}. If no pair of vertices in $\{v_1,v_2,v_3\}$
is a close or similar pair, we call $\{v_1,v_2,v_3\}$ a \emph{proper triple} of degree-($d+1$) vertices.
Our idea is to avoid bad cases and only branch on proper triples.

\begin{figure}
  \centering
  \includegraphics[width=0.8\textwidth]{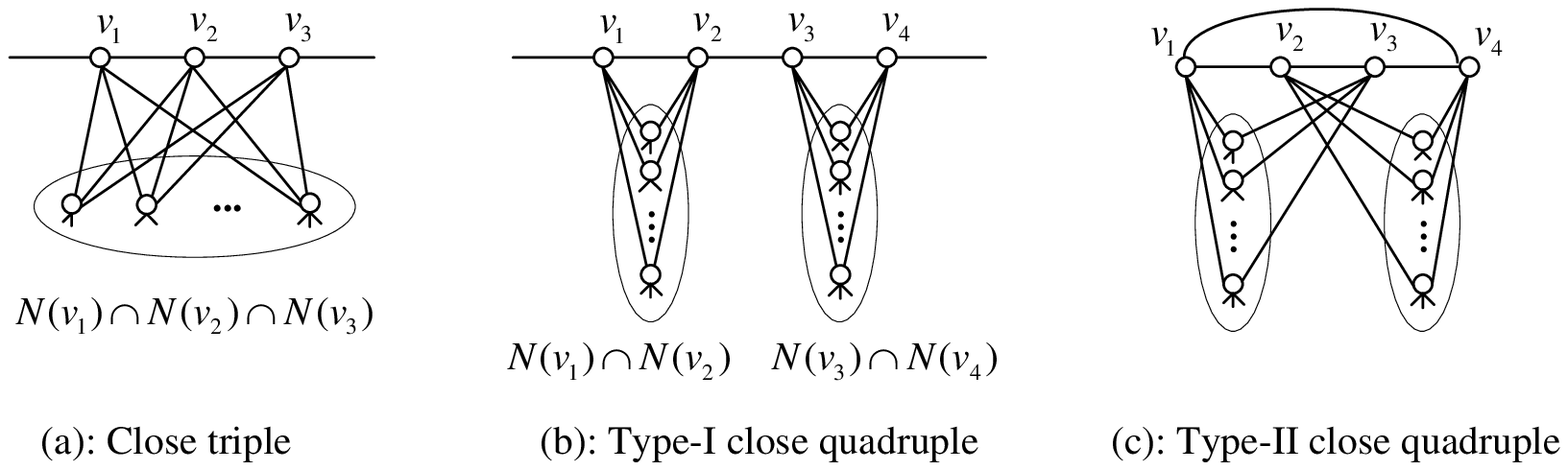}\\
  \caption{Illustrations of some structures}\label{three}
\end{figure}

Consider  four degree-($d+1$) vertices  $v_1, v_2, v_3$ and $v_4$ such that there is an edge between $v_i$ and $v_{i+1}$ for $i=1,2,3$.
If at most one pair of vertices in $\{v_1,v_2,v_3, v_4\}$
is a close or similar pair, then at least one of $\{v_1,v_2,v_3\}$ and $\{v_2,v_3, v_4\}$ will be a proper triple.
Thus the only left cases are that at least two pairs of vertices in $\{v_1,v_2,v_3, v_4\}$
are close or similar pairs.
Luckily, we find good branching rules to deal with them.
When both of $\{v_1,v_2\}$ and $\{v_2, v_3\}$
are close pairs, $\{v_1,v_2,v_3\}$ is called a \emph{close triple}. See Figure~\ref{three}(a) for an illustration of close triple. Our algorithm deals with close triples in Step 4 in Section~\ref{sec_alg}.
When both of $\{v_1,v_2\}$ and $\{v_3, v_4\}$
are close pairs, $\{v_1,v_2,v_3,v_4\}$ is called a \emph{type-I close quadruple}. See Figure~\ref{three}(b) for an illustration of type-I close quadruple. Our algorithm deals with type-I close quadruples in Step 5 in Section~\ref{sec_alg}.
When both of $\{v_1,v_3\}$ and $\{v_2, v_4\}$
are similar pairs, $\{v_1,v_2,v_3,v_4\}$ is called a \emph{type-II close quadruple}. See Figure~\ref{three}(c) for an illustration of type-II close quadruple. Our algorithm deals with type-II close quadruples in Step 6 in Section~\ref{sec_alg}.
When $\{v_1,v_2,v_3,v_4\}$ has one close pair and one similar pair, we can see that there is always a close triple in it.
Therefore, we have considered all possible cases.
The last step of our algorithm is then to deal with proper triples.

\section{The algorithm and its analysis}\label{sec_alg}
%First of all, we design an algorithm $BDD(G,d,k)$, which works for any $d\geq 0$. However, this algorithm can only achieve
%the running time bound of $O()$ for each $d\geq 3$. We will modify some steps of $BDD(G,d,k)$ to get an improved algorithm for the case that $d=2$ and
%design a totally new algorithm for the case that $d=1$.

We are ready to describe the whole algorithm. Our algorithm works for any $d\geq 0$ but can only achieve
the running time bound of $O^*((d+1)^k)$ for each $d\geq 3$.
Our algorithm is a recursive algorithm containing seven major steps, each of which will branch on the current instance into several sub-instances and invoke the algorithm itself on each sub-instance. Next, we describe these steps. When we introduce one step, we assume that all pervious steps can not be applied anymore.
For the purpose of presentation, we will analyze the correctness and running time of each step after describing it.

\medskip
\noindent \textbf{Step~1} (\textbf{Vertices of degree $\geq d+2$})

If there is a vertex $v$ of degree $\geq d+2$ in the graph, we branch on $v$ into $1+ {{d(v)}\choose{d(v)-d}}$
branches according to \refl{base1} by either  including $v$ or each set of $d(v)-d$ neighbors of $v$ to the solution set.

%The correctness of this step is based on the observation that in a yes-instance, if a vertex $v$ of degree $\geq d+2$ is not in the solution set, then at least
%$d(v)-d$ neighbors of $v$ are in the solution set.

In the branch where $v$ is included to the solution set, we delete $v$ from the graph and decrease the parameter $k$ by 1.
In the branch where a set $N'\subseteq N(V)$ of $d(v)-d$ neighbors of $v$ are included to the solution set, we delete $N'$ from
the graph and decrease the parameter $k$ by $d(v)-d$. For this operation, we get a recurrence relation
\eqn{e_step1}{
C(k)\leq C(k-1) + {{d(v)}\choose{d(v)-d}}\cdot C(k-(d(v)-d)).
}
Let $\gamma$ denote the branching factor of \refe{e_step1}.

\lem{l1}{If $d(v)-d\geq 2$, the branching factor $\gamma$ of \refe{e_step1} satisfies that
\eqn{e_le1}
{\gamma \leq \frac{{1 + \sqrt {2{d^2} + 6d + 5} }}{2}.}
}

A proof of this lemma can be found the full version of this paper. It is easy to verify that $\gamma \leq d+1$ for $d\geq 2$.
After Step~1, the graph has maximum degree $d+1$.

\medskip
\noindent \textbf{Step~2} (\textbf{Proper dominations})

If a vertex $v$ of degree $d+1$  is dominated by a vertex $u$ (or dominates a neighbor $u$ of it),
we branch on $v$ into $d(v)$
branches by including each vertex in  $N(v)$ (or $N[v]\setminus \{u\}$) to the solution set.
The correctness of this step is based on \refl{domination} and \refl{domination2}.

In each branch, a vertex is included to the solution set and $k$ decreases by 1.
Vertex $v$ is of degree $d+1$ since the graph has maximum degree at most $d+1$ after Step~1.
We get a recurrence relation
%\eqn{e_step2}{}
\[C(k)\leq d(v)\cdot C(k-1)= (d+1)\cdot C(k-1),
\]
the branching factor of which is $d+1$.

\medskip
\noindent \textbf{Step~3} (\textbf{Good pairs of degree-($d+1$) vertices})

Recall that a pair of adjacent degree-($d+1$) vertices is a \emph{good pair} if they have at least one
and at most $d-2$ common neighbors.
we use the following branching rule to deal with a good pair $\{v_1,v_2\}$.
Let $N^+=(N(v_1)\cap N(v_2))\cup \{v_1,v_2\}$, $N_1=N(v_1)\setminus N^+$ and $N_2=N(v_2)\setminus N^+$.
Assume that $v_1$ and $v_2$ have $x$ common neighbors.
Note that for any  $d$-degree deletion set $S'$, if $S'$ does not contain any vertex in $N^+$, then $S'$
contains at least one vertex in $N_1$ and one vertex in $N_2$.
We branch into $|N^+|+|N_1||N_2|=(x+2)+(d-x)^2$ branches. In the first $|N^+|$ branches each vertex in  $N^+$
is included to the solution set; and in the last $|N_1||N_2|$ branches each pair of vertices in $N_1$ and $N_2$ is
included to the solution set. In each branch, if $z$ vertices are included to the solution set, then the parameter $k$ in this branch decreases by $z$.
 This branching operation gives a recurrence relation
\[
C(k) \leq (x+2)\cdot C(k-1)+ (d-x)^2\cdot C(k-2),
\]
the branching factor of which is
\[\frac{1}{2}\left( {2 + x + \sqrt {5{x^2} - 8dx + 4{d^2} + 4x + 4} } \right).\]
It is easy to verify that when $1\leq x \leq d-2$, the branching factor is at most $d+1$.

\medskip
\noindent \textbf{Step~4} (\textbf{Close triples of degree-($d+1$) vertices})

Recall that a pair of adjacent degree-($d+1$) vertices is a \emph{close pair} if they have exactly $d-1$ common neighbors.
The formal definition of close triple is that: the set of three degree-($d+1$) vertices $v_1, v_2$ and $v_3$ is called a \emph{close triple} if $\{v_1,v_2\}$ and $\{v_2, v_3\}$
are two close pairs and $v_1$ and $v_3$ are not adjacent.
According to the definition of close triples, we can see that $N(v_1)\cap N(v_2)\cap N(v_3)=N(v_2)\setminus\{v_1,v_3\}$.
For a close triple $\{v_1,v_2,v_3\}$, we observe the following.
Vertex $v_1$ (resp., $v_3$) is adjacent to a degree -$(d+1)$ vertex $v_0\not\in N[v_2]$ (resp., $v_4\not\in N[v_2]$) by \refl{domi3}.
Let $N^-_2=N[v_2]\setminus\{v_1,v_3\}$.
For any $d$-degree deletion set $S'$, if $S\cap N^-_2=\emptyset$, then $S'$ contains either $v_1$ and a vertex in $\{v_3,v_4\}$ (since $S'$ must contain a vertex in $N[v_2]$ and a vertex in $N[v_3]$) or $v_3$ and a vertex in
$\{v_0,v_1\}$ (since $S'$ must contain a vertex in $N[v_2]$ and a vertex in $N[v_1]$). Then we can branch
by either including each vertex in $N^-_2$ to the solution set or including each of $\{v_1,v_3\}$, $\{v_1,v_4\}$
and $\{v_0,v_3\}$ to the solution set. This branching operation gives a recurrence relation
\[
C(k) \leq (d-1)\cdot C(k-1)+ 3\cdot C(k-2),
\]
the branching factor of which is
\[\frac{1}{2}\left( {d-1 + \sqrt {{d^2} - 4d + 13} } \right).\]
It is easy to verify that when $d\geq 2$, the branching factor is less than $d+1$.

\medskip
\noindent \textbf{Step~5} (\textbf{Type-I close quadruples of degree-($d+1$) vertices})

A set of four degree-($d+1$) vertices $\{v_1,v_2,v_3,v_4\}$ is called a \emph{type-I close quadruple} if
 $\{v_1,v_2,v_3,v_4\}$ induces a cycle or a path of 4 vertices, and
 $\{v_1,v_2\}$ and $\{v_3,v_4\}$ are two close pairs.
Let $N_{12}^-=N(v_1)\cap N(v_2)$ and $N_{34}^-=N(v_3)\cap N(v_4)$.
When the graph has no proper dominations, good pairs or close triples, it holds that
$N_{12}^- \cap N_{34}^-=\emptyset$.

Let $S'$ be an arbitrary $d$-degree deletion set.
 Our branching rule for type-I close quadruples is different for the cases whether  $\{v_1,v_2,v_3,v_4\}$ induces  a cycle or a path.

\textbf{Case 1.} $\{v_1,v_2,v_3,v_4\}$ induces a cycle of 4 vertices:
We consider the following different subcases.

\textbf{Case 1.1.}  $S'\cap \{v_1,v_2,v_3,v_4\}=\emptyset$: Then $S'\cap N_{12}^-\neq \emptyset$ and $S'\cap N_{34}^-\neq \emptyset$.
For this case, we included each pair of vertices in $N_{12}^-$ and $N_{34}^-$ to the solution set to create $|N_{12}^-||N_{34}^-|=(d-1)^2$ branches, each of which decreases $k$ by 2.

\textbf{Case 1.2.} $S'\cap \{v_1,v_2,v_3,v_4\}=\{v_1\}$ or $S'\cap \{v_1,v_2,v_3,v_4\}=\{v_2\}$: Then $S'\cap  N_{34}^-\neq \emptyset$,
otherwise no vertex in $N[v_3]$ or $N[v_4]$  would be in $S'$ and then $S'$ would not be a $d$-degree deletion set.
Furthermore, if $S'\cap \{v_1,v_2,v_3,v_4\}=\{v_2\}$, then $S'\setminus \{v_2\}\cup\{v_1\}$ is still a $d$-degree deletion set of the same size, since $N[v_2]\setminus N[v_1]=\{v_3\}$, $v_3$ is adjacent to all vertices in $N_{34}^-$
and $S'\cap  N_{34}^-\neq \emptyset$.
So for this case, we include $\{v_1,x\}$ to  the solution set for each $x\in N_{34}^-$ to create $|N_{34}^-|=d-1$ branches, each of which decreases $k$ by 2.

\textbf{Case 1.3.} $S'\cap \{v_1,v_2,v_3,v_4\}=\{v_3\}$ or $S'\cap \{v_1,v_2,v_3,v_4\}=\{v_4\}$: Then $S'\cap  N_{12}^-\neq \emptyset$. For the same reason, we
include $\{v_3,x\}$ to  the solution set for each $x\in N_{12}^-$ to create $|N_{12}^-|=d-1$ branches, each of which decreases $k$ by 2.

\textbf{Case 1.4.} $|S'\cap \{v_1,v_2,v_3,v_4\}|\geq 2$: Then $S'\setminus \{v_1,v_2,v_3,v_4\}\cup \{v_1,v_3\}$ is a $d$-degree deletion set of size not greater than that of $S'$, since $N[\{v_1,v_2,v_3,v_4\}]\subseteq N[\{v_1,v_3\}]$.
For this case, we can simply include $\{v_1,v_3\}$ to the solution set.

The branching operation gives a recurrence relation
\eqn{4-1-1}{
\begin{array}{*{20}{l}}
{C(k)}& \le &{{{(d - 1)}^2}\cdot C(k - 2)}+ (d - 1)\cdot C(k - 2) + (d - 1)\cdot C(k - 2) +C(k - 2)\\
{}& = &{d^2\cdot C(k - 2),}
\end{array}}
%\[
%C(k) \leq (d-1)^2C(k-2)+(d-1)C(k-2)+(d-1)C(k-2)+C(k-2)=d^2C(k-2),
%\]
the branching factor of which is $d<d+1$.

\textbf{Case 2.} $\{v_1,v_2,v_3,v_4\}$ induces a path of 4 vertices:
Let $\{v_0\}=N(v_1)\setminus N[v_2]$ and $\{v_5\}=N(v_4)\setminus N[v_3]$, where it is possible that $v_0=v_5$.
We observe the following different cases.

\textbf{Case 2.1.} $S'$ does not contain any vertex in $N_{12}^- \cup N_{34}^-$:
Then $S'$ contains at least one vertex in $\{v_0,v_1,v_2\}$ and at least one vertex in $\{v_3,v_4,v_5\}$,
since $S'$ must contain at least one vertex in $N[v_1]$ and at least one vertex in $N[v_4]$.
If $|S'\cap \{v_1,v_2,v_3,v_4\}|\geq 2$, then $S''=S'\setminus \{v_1,v_2,v_3,v_4\} \cup\{v_1,v_4\}$ is still a $d$-degree deletion set with $|S''|\leq |S'|$, since
$N[\{v_1,v_2,v_3,v_4\}]\subseteq N[\{v_1,v_4\}]$. Otherwise, it holds either $S'\cap \{v_0,v_1,v_2\}=\{v_0\}$ or
 $S'\cap \{v_3,v_4,v_5\}=\{v_5\}$.
If $S'\cap \{v_0,v_1,v_2\}=\{v_0\}$, then $v_3\in S'$ since $S'$ must contain at least one vertex in $N[v_2]$.
If $S'\cap \{v_3,v_4,v_5\}=\{v_5\}$, then $v_2\in S'$ since $S'$ must contain at least one vertex in $N[v_3]$.
So for this case, we conclude that there is a solution contains
one of $\{v_1,v_4\}$, $\{v_0,v_3\}$ and $\{v_2,v_5\}$.
In our algorithm, we generate three branches by including each of $\{v_1,v_4\}$, $\{v_0,v_3\}$ and  $\{v_2,v_5\}$ to the solution set. In each of the three branches, the parameter $k$ decreases by $2$.

\textbf{Case 2.2.} $S'$ does not contain any vertex in $N_{12}^-$ but contain some vertex in $N_{34}^-$:
Since $S'\cap N[v_1]\neq \emptyset$, we know that $S'$ contains at least one vertex in $\{v_0,v_1,v_2\}$.
If $v_2\in S'$, then $S''=S'\setminus \{v_2\} \cup\{v_1\}$ is still a $d$-degree deletion set. The reason relies on that
$N[v_2]\setminus N[v_1]=\{v_3\}$, $v_3$ is adjacent to each vertex in $N_{34}^-$, and $S''$
contains at least one vertex in $N_{34}^-$. So for this case, there is a solution contains one vertex in $\{v_0,v_1\}$.
In our algorithm, we create $2|N_{34}^-|=2(d-1)$ branches by including to the solution each pair of vertices $x$ and $y$ such that $x\in\{v_0,v_1\}$ and
$y\in N_{34}^-$. In each of the $2(d-1)$ branches, the parameter $k$ decreases by $2$.

\textbf{Case 2.3.} $S'$ does not contain any vertex in $N_{34}^-$ but contain some vertex in $N_{12}^-$:
For the same reason in Case~2.2, there is a  solution contains one vertex in $\{v_4,v_5\}$.
In our algorithm, we create $2|N_{12}^-|=2(d-1)$ branches by including to the solution each pair of vertices $x$ and $y$ such that $x\in\{v_4,v_5\}$ and
$y\in N_{12}^-$. In each of the $2(d-1)$ branches, the parameter $k$ decreases by $2$.

\textbf{Case 2.4.} $S'$ contains some vertex in $N_{12}^-$ and some vertex in $N_{34}^-$: For this case,
Our algorithm simply generates $|N_{12}^-||N_{34}^-|=(d-1)^2$ branches by including to the solution each pair of vertices $x$ and $y$ such that $x\in N_{12}^-$ and
$y\in N_{34}^-$. In each of the $(d-1)^2$ branches, the parameter $k$ decreases by $2$.

The above branching operation gives a recurrence relation

\[\begin{array}{*{20}{l}}
{C(k)}& \le &{3C(k - 2) + 2(d - 1)\cdot C(k - 2) + 2(d - 1)\cdot C(k - 2) + {{(d - 1)}^2}\cdot C(k - 2)}\\
{}& = &{d(d + 2)\cdot C(k - 2),}
\end{array}\]
the branching factor of which is $\sqrt {d(d+2)}< d+1$.

\medskip
\noindent \textbf{Step~6} (\textbf{Type-II close quadruples of degree-($d+1$) vertices})
%(\textbf{4-cycles with similar vertices}).

Two nonadjacent degree-($d+1$) vertices are \emph{similar} if they have the same neighbor set.
A set of four degree-($d+1$) vertices $\{v_1,v_2,v_3,v_4\}$ is called a \emph{type-II close quadruple}
%there is an edge between $v_i$ and $v_{i+1}$ for $i=1,2,3,4$, where $v_5=v_1$,
%$\{v_1,v_2,v_3,v_4\}$ induces either a cycle or path of 4 vertices,
%$\{v_1,v_3\}$ and $\{v_2,v_4\}$ are two similar pairs.
if $\{v_1,v_3\}$ and $\{v_2,v_4\}$ are two similar pairs and there is an edge between $v_i$ and $v_{i+1}$ for $i=1,2,3$.
Note that there must be an edge between $v_1$ and $v_4$ since $\{v_1,v_3\}$ is a similar pair.
So as a type-II close quadruple, $\{v_1,v_2,v_3,v_4\}$ always induces a cycle of 4 vertices.

Let $\{v_1,v_2,v_3,v_4\}$ be a  type-II close quadruple. We use $N_{13}^-$ to denote $N(v_1)\setminus\{v_2,v_4\}$ and
$N_{24}^-$ to denote $N(v_2)\setminus\{v_1,v_3\}$.
Note that it holds $N_{13}^-\cap N_{24}^-=\emptyset$, if we assume that there is no good pairs or close triples.
Let $S'$ be a  $d$-degree deletion set.
%Our branching rule for type-II close quadruples is similar to this for type-I close quadruples.
%different for the cases that $\{v_1,v_2,v_3,v_4\}$ induces  a cycle and a path.
%
%\textbf{Case 1.} $\{v_1,v_2,v_3,v_4\}$ induces a cycle of 4 vertices:
We consider the following different subcases.

\textbf{Case 1.}  $S'\cap \{v_1,v_2,v_3,v_4\}=\emptyset$: Then $S'\cap N_{13}^-\neq \emptyset$ and $S'\cap N_{24}^-\neq \emptyset$.
For this case, we included each pair of vertices in $N_{13}^-$ and $N_{24}^-$ to the solution set to create $|N_{13}^-||N_{24}^-|=(d-1)^2$ branches, each of which decreases $k$ by 2.

\textbf{Case 2.} $S'\cap \{v_1,v_2,v_3,v_4\}=\{v_1\}$ or $S'\cap \{v_1,v_2,v_3,v_4\}=\{v_3\}$: Then $S'\cap  N_{13}^-\neq \emptyset$,
otherwise $S'$ would not be a $d$-degree deletion set since no vertex in $N[v_3]$ or $N[v_1]$  is in $S'$.
Furthermore, if $S'\cap \{v_1,v_2,v_3,v_4\}=\{v_3\}$, then $S'\setminus \{v_3\}\cup\{v_1\}$ is still a $d$-degree deletion set of the same size.
So for this case, we include $\{v_1,x\}$ to  the solution set for each $x\in N_{13}^-$ to create $|N_{13}^-|=d-1$ branches, each of which decreases $k$ by 2.

\textbf{Case 3.} $S'\cap \{v_1,v_2,v_3,v_4\}=\{v_2\}$ or $S'\cap \{v_1,v_2,v_3,v_4\}=\{v_4\}$: Then $S'\cap  N_{24}^-\neq \emptyset$. For the same reason, we
include $\{v_2,x\}$ to  the solution set for each $x\in N_{24}^-$ to create $|N_{24}^-|=d-1$ branches, each of which decreases $k$ by 2.

\textbf{Case 4.} $|S'\cap \{v_1,v_2,v_3,v_4\}|\geq 2$: Then $S'\setminus \{v_1,v_2,v_3,v_4\}\cup \{v_1,v_2\}$ is a $d$-degree deletion set of size not greater than $S'$, since $N[\{v_1,v_2,v_3,v_4\}]\subseteq N[\{v_1,v_2\}]$.
For this case, we can simply include $\{v_1,v_2\}$ to the solution set.

The branching operation gives a recurrence relation
\[\begin{array}{*{20}{l}}
{C(k)}& \le &{{{(d - 1)}^2}\cdot C(k - 2)}+ (d - 1)\cdot C(k - 2) + (d - 1)\cdot C(k - 2) +C(k - 2)\\
{}& = &{d^2\cdot C(k - 2),}
\end{array}\]
%\[
%C(k) \leq (d-1)^2C(k-2)+(d-1)C(k-2)+(d-1)C(k-2)+C(k-2)=d^2C(k-2),
%\]
the branching factor of which is $d<d+1$.

\medskip
\noindent \textbf{Step~7} (\textbf{Proper triples of degree-($d+1$) vertices})

A set of three degree-($d+1$) vertices $\{v_1,v_2,v_3\}$ is called a \emph{proper triple} if
$\{v_1,v_2,v_3\}$ induces a path and no pair of vertices in $\{v_1,v_2,v_3\}$ is close or similar.

\lem{triple}{Let $G$ be a graph of maximum degree $d+1$ for any integer $d>0$. If $G$ has no proper dominations, good pairs, close triples, type-I close quadruples or type-II close quadruples,
then $G$ has some proper triples.
}
A proof of this lemma can be found in the full version.

\medskip

For a proper triple $\{v_1,v_2,v_3\}$  in a graph having none of dominated vertices, good pairs, close triples, type-I  close quadruples and type-II close quadruples, we have the following properties:
$N(v_1)\cap N(v_2)=\emptyset$, $N(v_2)\cap N(v_3)=\emptyset$ and $1\leq |N(v_1)\cap N(v_3)|\leq d$.

Let $N_{13}^-=N(v_1)\cap N(v_3)\setminus \{v_2\}$, $N_1^-=N(v_1)\setminus N(v_3)$, $N_3^-=N(v_3)\setminus N(v_1)$, $N_2^-=N(v_2)\setminus\{v_1,v_3\}$ and $x=|N_{13}^-|$.
Since $\{v_1,v_3\}$ is not a similar pair, we know that $0\leq x \leq d-1$.
Let $S'$ be a $d$-deletion set. To design our branching rule, we consider the following different cases.

\textbf{Case 1.} $v_2\in S'$: We simply include $v_2$ to the solution set and the parameter $k$ decreases by 1. For all the remaining cases, we assume that $v_2\not\in S'$.

\textbf{Case 2.} $v_2\not\in S'$ and $v_1,v_3\in S'$:  We simply include $v_1$ and $v_3$ to the solution set and the parameter $k$ decreases by 2.

\textbf{Case 3.} $v_1, v_2\not\in S'$ and $v_3\in S'$: For the case, $S'\cap (N(v_1)\setminus \{v_2\}\neq \emptyset$. We create  $|N(v_1)\setminus \{v_2\}|=d$ branches by including $v_3$ and each vertex
in $N(v_1)\setminus \{v_2\}$ to the solution set and the parameter $k$ in each branch decreases by 2.

\textbf{Case 4.} $v_2, v_3\not\in S'$ and $v_1\in S'$: For the case, $S'\cap (N(v_3)\setminus \{v_2\}\neq \emptyset$. We create  $|N(v_3)\setminus \{v_2\}|=d$ branches by including $v_1$ and each vertex
in $N(v_3)\setminus \{v_2\}$ to the solution set and the parameter $k$ in each branch decreases by 2.

\textbf{Case 5.} $v_1,v_2, v_3\not\in S'$: Then $S'$ must contains (i) a vertex in  $N_2^-$ and (ii)
either a vertex in $N_{13}^-$ or two vertices from $N_1^-$ and $N_3^-$ respectively.
Our algorithm generates $|N_2^-||N_{13}^-|+|N_2^-|||N_1^-||N_3^-|=(d-1)x+(d-1)(d-x)^2$ branches.
%In the first three branches, each vertex in $\{v_1,v_2,v_3\}$ is included to the solution set and the parameter $k$ decreases by 1.
Each of the first $(d-1)x$ branches includes a vertex in  $N_2^-$ and a vertex in $N_{13}^-$ to the solution set and the parameter $k$ decreases by 2.
The last  $(d-1)(d-x)^2$ branches are generated by including each triple $\{w_1\in N_2^-,w_2 \in N_1^-,w_3\in N_3^-\}$ to the solution set, where the parameter $k$ decreases by 3.

The above branching operation gives a recurrence relation
\eqn{wer}{
\begin{array}{*{20}{l}}
{C(k)}& \le &{C(k - 1) + C(k-2)+d\cdot C(k - 2) + d\cdot C(k - 2) +} \\
{}&  & (d - 1)x\cdot C(k - 2) + (d - 1)(d-x)^2\cdot C(k - 3)\\
{}& = &{ C(k-1)+  ((2d+1)+(d - 1)x)\cdot C(k - 2) + (d - 1)(d-x)^2\cdot C(k - 3),}
\end{array}}
%\[ C(k)\leq 3C(k-1)+(d-1)x\cdot C(k-2)+(d-1)(d-x)^2\cdot C(k-3),\]
where $0\leq x \leq d-1$.

\lem{factor2}{When $d\geq 3$, the branching factor of \refe{wer} is at most $d+1$ for each $0\leq x \leq d-1$.
}
A proof of this lemma can be found in the full version.

\subsection{The results}
\refl{triple} guarantees that when the graph has a vertex of degree $\geq d+1$, one of the above seven steps can be applied.
When $d\geq 3$, the branching factor in each of the seven steps is at most $d+1$. Thus,
\thm{result}{The $d$-bounded-degree vertex deletion problem for each $d\geq 3$ can be solved in $O^*((d+1)^k)$ time.}

Note that all the seven steps of our algorithm work for $d=2$. In the first six steps, we still can get branching factors at most $d+1$ for $d=2$.
In Step~7, when $d=2$ and $x=d-1=1$, \refe{wer} becomes
\[ C(k)\leq C(k-1)+6C(k-2)+C(k-3),\]
which has a branching factor of 3.0645. This is the biggest branching factor in the algorithm. Then
\thm{result}{The  co-path/cycle problem can be solved in $O^*(3.0645^k)$ time.}
Note that previously the co-path/cycle problem could only be solved deterministically in $O^*(3.24^k)$ time~\cite{CF:copath}.

\section{Concluding remarks}
In this paper, by studying the structural properties of graphs, we show that the $d$-bounded-degree vertex deletion problem
can be solved in $O^*((d+1)^k)$ time for each $d\geq 3$. Our algorithm is the first nontrivial parameterized algorithm for
the $d$-bounded-degree vertex deletion problem with $d\geq 3$.

Our problem is a special case of the $(d+2)$-hitting set problem. It is still left as an open problem that whether the  $d'$-hitting set problem
can be solved in $O^*((d'-1)^k)$ time. Our result is a step toward to this interesting open problem. However, our method can not be extended to the $d'$-hitting set problem directly, since
some good graph structural properties do not hold in the general $d'$-hitting set problem.

%\appendix

\end{document}